\documentclass[amsmath,amssymb,12pt]{article}
\usepackage{jheppub}
\pdfoutput=1
\usepackage[utf8]{inputenc}
\usepackage{amsfonts}
\usepackage[table]{xcolor}
\usepackage{graphicx}
\usepackage{slashed}
\usepackage{xcolor}
\usepackage{tensor}
\usepackage{bbm}
\usepackage{subcaption}
\usepackage{ulem}
\usepackage{longtable}
\usepackage{afterpage}
\usepackage{ulem}

\newcommand{\AAm}[1]{{#1}} 
\newcommand{\DKB}[1]{{#1}} 

\preprint{IFT-UAM/CSIC-21-1}

\title{Hydrodynamic magneto-transport in charge density wave states}
\author[1,2]{Andrea Amoretti,}
\author[3,4]{Daniel Are\'an,}
\author[1]{Daniel K. Brattan,} 
\author[1,2]{Nicodemo Magnoli.}

\emailAdd{andrea.amoretti@ge.infn.it}
\emailAdd{daniel.arean@uam.es}
\emailAdd{danny.brattan@gmail.com}
\emailAdd{nicodemo.magnoli@ge.infn.it}

\affiliation[1]{Dipartimento di Fisica, Universit\`a di Genova,
via Dodecaneso 33, I-16146, Genova, Italy}
\affiliation[2]{I.N.F.N. - Sezione di Genova, via Dodecaneso 33, I-16146, Genova, Italy}
\affiliation[3]{Departamento de F\'isica Te\'orica, Universidad Aut{\'o}noma de Madrid, Campus de Cantoblanco, 28049 Madrid, Spain}
\affiliation[4]{Instituto de F\'\i sica Te\'orica UAM/CSIC, Calle Nicol\'as Cabrera 13-15, 28049 Madrid, Spain}

\abstract{In this paper we study the dynamical properties of charged systems immersed in an external magnetic field and perturbed by a set of scalar operators breaking translations either spontaneously or pseudo-spontaneously. By combining hydrodynamic and quantum field theory arguments we provide analytic expressions for all the hydrodynamic transport coefficients relevant for the diffusive regime in terms of thermodynamic quantities and DC thermo-electric conductivities.  This includes the momentum dissipation rate. We shed light on the role of the momentum dissipation rate in the transition between the pseudo-spontaneous and the purely explicit regimes in this class of systems. Finally, we clarify several relations between the hydrodynamic transport coefficients which have been observed in the holographic literature of charge density wave models.}

\begin{document}

\maketitle

\section{Introduction}
\label{sec:intro}
{\ The stability of many time evolving processes relies on dissipation precluding a run-away behavior. In particular, momentum dissipation plays a key role in systems such as disordered materials or phases with pseudo-spontaneous spatial ordering. The analysis of the transport properties in systems at non-zero charge densities where translations are broken either spontaneously or pseudo-spontaneously, commonly referred as charge density wave systems, can be traced back to \cite{PhysRevB.17.535,PhysRevB.19.3970,RevModPhys.60.1129}. Particularly relevant in order to make contact with experimental setups is the study of such systems in the presence of an external magnetic field \cite{FUKUYAMA19751323,PhysRevB.18.6245}.  In recent times, the discovery of this type of ordering in strongly coupled materials such as high temperature superconductors (see \textit{e.g.}~\cite{traquada,PhysRevLett.88.167008}) has caused a revival of the experimental and theoretical approaches to the topic of pseudo-spontaneous spatial ordering. In particular, hydrodynamic descriptions of these systems both at zero~\cite{Delacretaz:2017zxd,Delacretaz:2016ivq} and non-zero~\cite{Delacretaz:2019wzh,Amoretti:2019buu}  external magnetic field have been developed in the last years, together with several holographic models~\cite{Andrade:2017cnc,Alberte:2017oqx,Amoretti:2017frz,Amoretti:2017axe,Alberte:2017cch,Amoretti:2018tzw,Andrade:2018gqk,Amoretti:2019kuf,Amoretti_2019,Ammon:2020xyv,Amoretti:2020ica,Andrade:2020hpu,Baggioli:2020nay,Baggioli:2020edn,Baggioli:2021xuv}.

{\ In this work we study $(2+1)$-dimensional systems at non-zero temperature ($T$) and chemical potential ($\mu$) in the presence of an external magnetic field ($B$), where \AAm{a set of scalar operators $O_I$, with $I=1,2$, account for the breaking of (spatial) translation invariance via a spatially modulated source $\Phi_I(x_i)$. This implies that the system will satisfy the following one-point function Ward identities
    \begin{eqnarray}
        \label{Eq:ConservationEquations2}
        \partial_{\mu} \langle T^{\mu \nu}\rangle = F^{\nu \mu}\langle  J_{\mu}\rangle
         - \left(\partial^{\nu} \Phi^{I}(x_i)\right) \langle O_{I}\rangle \; , \qquad \partial_{\mu} \langle J^{\mu}\rangle
         = 0 \, , 
    \end{eqnarray}
where $T^{\mu \nu}$ is the stress tensor, $F^{\mu \nu}$ an external electromagentic field strength, and $J^{\mu}$ a $U(1)$ charge current. More specifically, in what follows we will restrict our attention to systems where the vev of the scalar operators is proportional to the corresponding spatial coordinates $\langle O_I \rangle= \delta_I^i x_i$ and the derivative of the source $\Phi_I$ is a constant, $\partial_i \Phi^I= \varphi \delta_i^I$. This corresponds to a system which breaks translations but preserves the homogeneity of the Ward identities \eqref{Eq:ConservationEquations2}. The symmetry pattern we have in mind is the one described for example in \cite{Nicolis:2013lma,Nicolis:2015sra}, where translational symmetry and shifts of the scalar operators $O_I$ by a constant are both broken but the diagonal group (shift+translation) remains unbroken, ensuring the homogeneity of the equations of motion. Systems with this symmetry breaking pattern have been intensively studied in the past few years and applied to the constructions of effective field theories for lattice phonons \cite{Leutwyler:1993gf,Nicolis:2017eqo}, classifications of solid state phases \cite{Nicolis:2015sra} and hydrodynamic \cite{Delacretaz:2017zxd,Delacretaz:2016ivq} and holographic \cite{Andrade:2017cnc,Alberte:2017oqx,Amoretti:2017frz,Amoretti:2017axe,Alberte:2017cch,Amoretti:2018tzw,Andrade:2018gqk,Amoretti:2019kuf,Amoretti_2019,Ammon:2020xyv,Amoretti:2020ica,Andrade:2020hpu,Baggioli:2020nay,Baggioli:2020edn,Baggioli:2021xuv} construction of charge density wave state effective field theories. Due to the discussion above, and to simplify notation, in what follows we will suppress the difference between indices labeling the scalar operators $I$ and the spatial indices $i$ and treat them as equivalent.

{\ Different regimes for these systems can be defined by the relative values of source and vev. Whenever the source $\varphi$ is vanishing, but the vev $\langle O_{i} \rangle$ is non-zero in the ground state, we say that translations are broken spontaneously. This should be compared against the explicit situation where the $O_{i}$ acquire a non-zero vev only because $\varphi$ is non-zero. Within the explicit regime we can further distinguish the pseudo-spontaneous and truly explicit cases. In the former $\varphi \ll |\partial_i \langle O_i \rangle|$. In this case, the low energy degrees of freedom are pseudo-Goldstone bosons which have a small, non-zero mass which is related to the ``pinning frequency'' \cite{Delacretaz:2017zxd} (denoted $\omega_{0}^{2}$). On the other hand, a truly explicit case occurs when $\varphi \gtrsim |\partial_i \langle O_i \rangle|$. This happens for example in the models of \cite{Andrade:2013gsa}, where only the source is non-zero and the vev is vanishing.}}

{\ In both the spontaneous and explicit regimes the conservation laws \eqref{Eq:ConservationEquations2} imply Ward identities involving the two point functions of the system. It has been known for a long time \cite{Hartnoll_ohm} that for charged fluids in an external magnetic field, the thermo-electric conductivities are inter-related in such a way that knowing the electric conductivity at non-zero frequency is sufficient to determine the thermo-electric and thermal conductivities. We shall demonstrate that this structure generalises to the systems described above and, as a consequence, we find non-trivial constraints on the low frequency AC conductivities.}

{\ The results from the Ward identities provide important constraints when we subsequently construct a complete description of the hydrodynamics of this class of models in the diffusive sector to first order in fluctuations and derivatives. We use the approach of \cite{Amoretti_2020} to determine analytically, to all orders in the magnetic field, the hydrodynamic transport coefficients which appear in the constitutive relations relevant for the diffusive sector (i.e.~the regime where perturbations decay without propagating in any particular spatial direction). This method \cite{Amoretti_2020} ensures that hydrodynamics always reproduces the DC conductivities of the system (including the thermal conductivities). The resulting analytic expressions are one of our key results and allow us to perform an analysis of the behaviour of these systems in various regimes without resorting to numerics.}

{\ One of the most important consequences of the Ward identities for hydrodynamics concerns the types of dissipation that can appear in the hydrodynamic equations of motion. Generically, when translation invariance is broken, one allows for explicit dissipation terms, \textit{i.e.}
    \begin{eqnarray}
        \partial_{t} P^{i} \sim - \Gamma^{ij} P_{j} + F^{ij} J_{j} - \left(\partial^{i} \Phi^{j}\right) O_{j} \, , \qquad \partial_{t} O^{i} \sim - \Omega^{ij} O_{j} \; ,
    \end{eqnarray}
where $P_{i}$ is the spatial momentum. The first equation contains the momentum dissipation tensor $\Gamma^{ij}$, whose precise hydrodynamic definition we shall give later. The second expression contains a phase relaxation tensor $\Omega^{ij}$ which accounts for the spreading out that occurs for a massive scalar field as it evolves (it is not protected by a conservation law). As a precise consequence of the Ward identities coming from \eqref{Eq:ConservationEquations2} we find that $\Gamma^{ij} = 0$. We shall demonstrate however the emergence of an effective $\Gamma^{ij}$, given in terms of the phase relaxation $\Omega^{ij}$ and the pinning frequency $\omega_0$, in regimes where the scalar field is effectively kinetic - by which we mean that the scalar does not evolve in time. This approximate independence from time evolution defines the strongly explicit regime.}

{\ With this result to hand, we are also in a position to clarify and extend a key observation of \cite{Amoretti:2018tzw}, further discussed in \cite{Donos:2019hpp,Baggioli:2020nay,Baggioli:2020haa}, namely that in the limit of small pinning frequencies the phase relaxation coefficients (longitudinal and Hall) are proportional to the square of the pinning frequency. We also suggest a way to identify the constant of proportionality to be the Goldstone susceptibilities \cite{Amoretti:2018tzw,Baggioli:2020nay,Baggioli:2020haa} at low pinning frequency, independently of the value of the magnetic field.}

{\ The outline of the paper is as follows: in section \ref{sec2} we discuss the consequences of the Ward identities for the correlators of the system we have
outlined above (without restricting ourselves to any particular regime). In section \ref{sec:hydrosec} we introduce the hydrodynamic description and compute the AC correlators. At the end of this section we derive novel and important results for the transport properties of the system that constitute the main outcome of this paper. Finally, in section \ref{sec:conclusions} we draw some conclusions and discuss future directions.}

\section{Ward identities from field theory}
\label{sec2}
{\ In the subsequent sections we will discuss the consequences of the Ward identities for AC transport in the diffusion sector. Hence, as our concern will primarily be in the regime described by diffusive hydrodynamics, we will deal with heat rather than spatial momentum ($P^{i}$) conservation. Therefore we define the spatial canonical heat current ($Q^{i}$) by
    \begin{eqnarray}
   \langle  Q^{i} \rangle = \langle P^{i}\rangle - \mu \langle J^{i} \rangle \; ,
    \end{eqnarray}
where $i$ indexes spatial directions. It is important to note that in the non-linear hydrodynamic theory the canonical heat current is generally not the one that enters into the entropy conservation equation; this will be relevant when we discuss the constitutive relations as the leading term in the spatial heat current will differ from $s T$.}

\subsection{Spontaneous case}

{\  \AAm{We wish to obtain the Ward identities that follow from \eqref{Eq:ConservationEquations2} when considering the consequences of broken translational symmetry (with zero scalar source $\varphi=0$) and unbroken $U(1)$ gauge invariance. As we stated above, we are restricting to cases in which the Ward identities are homogeneous. This will allow us to Fourier transform them, following the convention where for a function $f(t,\vec{x})$,
    \begin{eqnarray}
     f(t,\vec{x}) = \int \frac{d^{2}k d\omega}{(2 \pi)^3} f(\omega,\vec{k}) e^{- i ( \omega t - i \vec{k} \cdot \vec{x} )} \; .
    \end{eqnarray}}
Consequently, at arbitrary frequency and zero wavevector, the desired Ward identities are
    \begin{subequations}
    \label{Eq:SpontaneousWard}
    \begin{eqnarray}
        \label{Eq:SpontaneousWard1}
        i \omega \langle Q^{i} Q^{j} \rangle &=& - \left( i \omega \mu \delta\indices{^{i}_{k}} - F\indices{^{i}_{k}} \right) \langle Q^{k} J^{j} \rangle - i \omega \left( \chi_{\pi \pi} - \mu n \right) \delta^{ij} \; , \\
        \label{Eq:SpontaneousWard2}
        i \omega \langle Q^{i} J^{j} \rangle &=& - \left( i \omega \mu \delta\indices{^{i}_{k}} - F\indices{^{i}_{k}} \right) \langle J^{k} J^{j} \rangle - i \omega n \delta^{ij} \; , \\
        F_{ij} &=& B \epsilon_{ij} \; , \qquad \epsilon_{12} = 1 \; , \nonumber
    \end{eqnarray}
    \end{subequations}
where $n$ is the electric charge density and $\chi_{\pi \pi}$ is the momentum susceptibility which the reader can either define through the contact term in the Ward identity \eqref{Eq:SpontaneousWard1} or through the thermodynamic susceptibility we discuss in section \ref{sec:hydrosec}. There is a third Ward identity involving the scalar which requires a little care, as the stress energy tensor in the presence of external sources must be carefully identified before varying it with respect to the scalar source. Following the procedure described in \cite{Bianchi:2001de,Amoretti:2016bxs,Amoretti:2020ica}, \AAm{and remembering that since $\langle O_i \rangle=x_i$ then $\partial_{i} \langle O_{j} \rangle = \delta_{ij}$},  we obtain:
    \begin{equation} \tag{2.4c}
        \label{Eq:SpontaneousWard3}
        i \omega \langle Q^{i} O^{j} \rangle = - \left( i \omega \mu \delta\indices{^{i}_{k}} - F\indices{^{i}_{k}} \right) \langle J^{k} O^{j} \rangle + \delta^{ij} \; .
    \end{equation}
From these three Ward identities, one can see that knowing the correlators $\langle J^{i} J^{j} \rangle$ and $\langle J^{i} O^{j} \rangle$ to arbitrary frequency is equivalent to knowing all of the correlators in \eqref{Eq:SpontaneousWard}. This follows from the ladder structure, \textit{e.g.}~knowing $\langle J^{i} O^{j} \rangle$ we can solve \eqref{Eq:SpontaneousWard3} for $\langle Q^{i} O^{j} \rangle$. Similarly, $\langle J^{i} J^{j} \rangle$ gives $\langle Q^{i} J^{j} \rangle$ through \eqref{Eq:SpontaneousWard2}, and subsequently $\langle Q^{i} Q^{j} \rangle$ through \eqref{Eq:SpontaneousWard1}.}

{\ We define the following AC transport quantities 
    \begin{eqnarray}
      \label{Eq:SpontaneousAC}
      (\sigma,\alpha,\kappa)^{ij}(\omega) = \frac{1}{i \omega} \left(\langle J^{i} J^{j} \rangle, \langle Q^{i} J^{j} \rangle, \langle Q^{i} Q^{j} \rangle \right) \; , \nonumber \\
      (\gamma,\theta)^{ij}(\omega) = (\langle J^{i} O^{j} \rangle, \langle Q^{i} O^{j} \rangle ) \; , \qquad
      \xi^{ij}(\omega) = i \omega \langle O^{i} O^{j} \rangle \; , 
    \end{eqnarray}
with the first three being the standard thermo-electric conductivities. There is an unfortunate but deliberate overlap between the typical notation for the transport coefficients that will appear in the hydrodynamic constitutive relations and the corresponding AC conductivities. We shall indicate the latter, and also their DC limits, with an explicit argument. The former will always appear without an argument. Using our assumption that translation invariance is broken homogeneously, the AC transport tensors of \eqref{Eq:SpontaneousAC} can be decomposed with respect to $SO(2)$ rotation invariance into longitudinal and Hall parts,
    \begin{eqnarray}
    \label{Eq:Spontaneousdecomp}
             (\sigma,\alpha,\kappa,\gamma,\theta,\xi)_{ij}(\omega) &=& (\sigma,\alpha,\kappa,\gamma,\theta,\xi)_{\mathrm{(L)}}(\omega) \delta_{ij}  
                                                                   - (\sigma,\alpha,\kappa,\gamma,\theta,\xi)_{\mathrm{(H)}}(\omega) (F^{-1})_{ij} \; , \qquad
    \end{eqnarray}
where we have employed the inverse of the field strength in our decomposition as this introduces an overall $1/B$ factor. We use the field strength $F_{ij}$ to decompose our tensors rather than the more typical two-dimensional Levi-Civita tensor $\epsilon_{ij}$ because we wish to emphasise that we are considering a theory that preserves spatial parity invariance microscopically. Any parity breaking is due to the presence of a fixed magnetic field $B$. The Hall conductivities $\sigma_{(\mathrm{H})}$ et cetera will be smooth in the vanishing magnetic field limit with our choice of decomposition \eqref{Eq:Spontaneousdecomp}. This should be compared to the explicit case \eqref{Eq:Explicitdecomp} where the limit as $B \rightarrow 0$ of the Hall conductivities vanishes and we decompose the ``Hall sector'' with respect to $F_{ij}$ instead.}

{\ Using \eqref{Eq:Spontaneousdecomp} and substituting into \eqref{Eq:SpontaneousWard} gives relations between the AC transport coefficients for arbitrary frequency. We can thus take $\sigma^{ij}(\omega)$ and $\gamma^{ij}(\omega)$ to be the ``independent'' transport terms; all others, with the special exception of $\xi^{ij}(\omega)$ which does not appear in the Ward identities, can be derived from them using \eqref{Eq:SpontaneousWard}. Of particular interest are the low frequency expansions of these independent terms ($\sigma^{ij}(\omega)$ and $\gamma^{ij}(\omega)$):
    \begin{subequations}
    \label{Eq:LowFrequencySpontaneousWard}
    \begin{eqnarray}
      \label{Eq:LowFrequencySpontaneousWard1}
      \sigma_{(\mathrm{L})}(\omega) &=& - \frac{i}{B^2} \left( \mu n - \alpha_{(\mathrm{H})}(0) \right) \omega + \frac{\kappa_{(\mathrm{L})}(0)}{B^2} \omega^2 + \mathcal{O}(\omega^3) \; , \\
      \sigma_{(\mathrm{H})}(\omega) &=& \sigma_{(\mathrm{H})}(0) + \frac{\kappa_{\mathrm{H}}(0) - \mu \left( 2 \chi_{\pi \pi}- \mu n \right)}{B^2} \omega^2 + \mathcal{O}(\omega^3) \; , \\
      \gamma_{(\mathrm{L})}(\omega) &=& \frac{i\left( \mu + \theta_{\mathrm{(H)}}(0) \right)}{B^2} \omega + \mathcal{O}(\omega^2) \; , \\
      \gamma_{(\mathrm{H})}(\omega) &=& \gamma_{(\mathrm{H})}(0) - i \theta_{\mathrm{(L)}}(0)\omega + \mathcal{O}(\omega^2) \; ,
    \end{eqnarray}
    \end{subequations}
where
    \begin{eqnarray}
     \label{Eq:Symmetrydictatedconductivities}
     \sigma_{(\mathrm{L})}(0) = \alpha_{(\mathrm{L})}(0) = \gamma_{(\mathrm{L})}(0) = 0 \; , \qquad \sigma_{(\mathrm{H})}(0) = - n \; , \nonumber \\
     \alpha_{(\mathrm{H})}(0) = \mu n - \chi_{\pi \pi}
     \; , \qquad \gamma_{(\mathrm{H})}(0) = 1 \; . \qquad
    \end{eqnarray}
The values obtained in \eqref{Eq:Symmetrydictatedconductivities} are a pure consequence of symmetry and the definition of the canonical heat; for any system regardless of its microscopic formulation that satisfies \eqref{Eq:SpontaneousWard}, the displayed DC conductivities will take this form. Further, as these expansions are purely consequences of the Ward identities, they must hold for arbitrary magnetic fields and temperatures.}

{\ It was noted in \cite{Amoretti_2020} that the thermo-electric and thermal conductivities are responsible for the $\mathcal{O}(\omega)$ and $\mathcal{O}(\omega^2)$ terms in the AC electric conductivity. As we have shown, this remains true even when spatial translation invariance is spontaneously broken by scalar operators \textit{e.g.}~if one knows for example $\sigma_{(\mathrm{L})}''(0)$, then one can determine $\kappa_{(\mathrm{L})}(0)$ from \eqref{Eq:LowFrequencySpontaneousWard1}. Moreover, the DC electric and thermo-electric conductivities are still constrained by symmetry, such that $\kappa_{(\mathrm{L})}(0)$ and $\kappa_{(\mathrm{H})}(0)$ are the system dependent DC conductivities in the suite of thermo-electric conductivities. The expressions for the AC longitudinal and Hall electric conductivities at low frequency \eqref{Eq:LowFrequencySpontaneousWard} are also the same as those in the absence of the spontaneously broken translation invariance (modulo replacement of the thermal conductivities by their system dependent values).}

{\ Additional correlators involving the scalar are also relevant to the low frequency behaviour of the system. As we can see from \eqref{Eq:LowFrequencySpontaneousWard} these correlators form a distinct set at low frequency. Symmetry dictates the value of the $\gamma_{(\mathrm{L})}(0)=0$ and $\gamma_{(\mathrm{H})}(0)=1$, see \eqref{Eq:Symmetrydictatedconductivities}, which are the zero frequency limits of the scalar-electric current correlators. This leaves $\theta_{(\mathrm{L})}(0)$ and $\theta_{(\mathrm{H})}(0)$ to carry the system dependent information which appears at $\mathcal{O}(\omega)$ in the low frequency expansions of $\gamma_{(\mathrm{L})}(\omega)$ and $\gamma_{(\mathrm{H})}(\omega)$. Completely in opposition to what we will find in the explicit case, the scalar-scalar correlator ($\xi^{ij}(\omega)$) plays no role in \eqref{Eq:LowFrequencySpontaneousWard}.}

\subsection{Explicit case}

{\ \AAm{In the explicit case, the translation breaking scalars have a non-zero source in the background $\varphi \neq 0$. The 2-pt Ward identities at arbitrary frequency and zero wavevector can be derived from \eqref{Eq:ConservationEquations2} in the same way as in the spontaneous case \cite{Amoretti:2016bxs}:}
    \begin{subequations}
        \label{Eq:ExplicitWardIdentity}
        \begin{eqnarray}
            \label{Eq:ExplicitWardIdentity1}
            i \omega \langle Q^{i} Q^{j} \rangle &=& - \left( i \omega \mu \delta\indices{^{i}_{k}} - F\indices{^{i}_{k}} \right) \langle Q^{k} J^{j} \rangle + \varphi \langle Q^{i} O^{j} \rangle 
            - i \omega \left( \chi_{\pi \pi} - \mu n \right) \delta^{ij} \; , \qquad \\
            \label{Eq:ExplicitWardIdentity2}
            i \omega \langle Q^{i} J^{j} \rangle &=& - \left( i \omega \mu \delta\indices{^{i}_{k}} - F\indices{^{i}_{k}} \right) \langle J^{k} J^{j} \rangle + \varphi\langle J^{i} O^{j} \rangle - i \omega n \delta^{ij} \; , \\
            \label{Eq:ExplicitWardIdentity3}
            i \omega \langle Q^{i} O^{j} \rangle &=& - \left( i \omega \mu \delta\indices{^{i}_{k}} - F\indices{^{i}_{k}} \right) \langle J^{k} O^{j} \rangle - \varphi\langle O^{i} O^{j} \rangle + \delta^{ij} \; ,
        \end{eqnarray}
    \end{subequations}
Unlike the spontaneous case, we can see that three correlators, $\langle J^{i} J^{j} \rangle$, $\langle J^{i} O^{j} \rangle$ and $\langle O^{i} O^{j} \rangle$ are required to specify the arbitrary frequency behaviour of the system; compared to just the first two for the spontaneous case as all other correlators can be derived from them through \eqref{Eq:ExplicitWardIdentity}. Clearly the scalar-scalar correlator will now play an integral role in DC transport.}

{\ We define the following new AC quantities 
    \begin{eqnarray}
      (\varpi,\vartheta)^{ij}(\omega) = \frac{1}{i \omega} (\langle J^{i} O^{j} \rangle, \langle Q^{i} O^{j} \rangle \rangle) \; , \qquad
      \zeta^{ij}(\omega) = \frac{1}{i \omega} \left( \langle O^{i} O^{j} \rangle - \frac{1}{\varphi} \delta^{ij} \right) \;  , \qquad
    \end{eqnarray}
which replace $\gamma^{ij}$, $\theta^{ij}$ and $\xi^{ij}$ in \eqref{Eq:SpontaneousAC}. Notice the introduction of an $i \omega$ factor into the definitions of $\varpi$ and $\vartheta$ in the explicit case in comparison to the spontaneous case. The definition of the standard thermo-electric conductivities are the same as in the spontaneous case. Moreover, we now decompose the tensor transport quantities with respect to $SO(2)$ rotational invariance according to
    \begin{eqnarray}
    \label{Eq:Explicitdecomp}
             (\sigma,\alpha,\kappa,\varpi,\vartheta,\zeta)_{ij}(\omega) &=& (\sigma,\alpha,\kappa,\varpi,\vartheta,\zeta)_{\mathrm{(L)}}(\omega) \delta_{ij}  
                                                                   + (\sigma,\alpha,\kappa,\varpi,\vartheta,\zeta)_{\mathrm{(H)}}(\omega) F_{ij} \; . \qquad
    \end{eqnarray}
This decomposition introduces an additional factor of $B^2$ into the Hall conductivities compared to \eqref{Eq:Spontaneousdecomp}. This has been done because in the explicit case the Hall conductivities are smooth in the $B \rightarrow 0$ limit.}

{\ Once again we may employ the Ward identities to determine the low frequency expansions of the independent AC transport terms. From \eqref{Eq:ExplicitWardIdentity2} we see that knowing $\sigma^{ij}(\omega)$ and $\varpi^{ij}(\omega)$ at arbitrary frequency, allows us to solve for $\alpha^{ij}(\omega)$. Similarly, from \eqref{Eq:ExplicitWardIdentity3}, given $\varpi^{ij}(\omega)$ and $\zeta^{ij}(\omega)$, we can obtain $\vartheta^{ij}(\omega)$. Finally, using these previously determined quantities we can solve for $\kappa^{ij}(\omega)$ through \eqref{Eq:ExplicitWardIdentity1}. Therefore, $\sigma^{ij}(\omega)$, $\varpi^{ij}(\omega)$ and $\zeta^{ij}(\omega)$ are the independent AC transport quantities, as everything else can be determined in terms of them, and they have the following form at low frequency
    \begin{subequations}
    \label{Eq:Explicitwardlowfrequency}
    \begin{eqnarray}
        \label{Eq:Explicitwardlowfrequency1}
        \sigma_{(\mathrm{L})}(\omega) &=& \frac{\varphi^2 \zeta_{(\mathrm{L})}(0)}{B^2} + \mathcal{O}(\omega) \; , \\
        \label{Eq:Explicitwardlowfrequency2}
        \sigma_{(\mathrm{H})}(\omega) &=& - \frac{n - \varphi^2 \zeta_{(\mathrm{H})}(0)}{B^2} + \mathcal{O}(\omega) \; , \\
        \label{Eq:Explicitwardlowfrequency3}
        \varpi_{(\mathrm{L})}(\omega) &=& \varphi \zeta_{(\mathrm{H})}(0) + \mathcal{O}(\omega) \; , \qquad \\
        \label{Eq:Explicitwardlowfrequency4}
        \varpi_{(\mathrm{H})}(\omega) &=& - \frac{\varphi \zeta_{(\mathrm{L})}(0)}{B^2} + \mathcal{O}(\omega) \; , \qquad
    \end{eqnarray}
    \end{subequations}
with the low frequency expansion of $\zeta^{ij}(\omega)$ unconstrained. Corrections up to and including $\mathcal{O}(\omega^2)$ of \eqref{Eq:Explicitwardlowfrequency1} and \eqref{Eq:Explicitwardlowfrequency2} involve at most the first and second derivatives of $\zeta^{ij}(\omega)$ evaluated at zero frequency, the DC thermal conductivities and $\varpi_{(\mathrm{L})}(\omega)$ and $\varpi_{(\mathrm{H})}(\omega)$ plus their first derivatives evaluated at zero frequency.}

{\ One can see from \eqref{Eq:Explicitwardlowfrequency} the crucial role played by the scalar operator in low frequency transport. It is entirely responsible for the now (\textit{c.f.}~the spontaneous case) non-zero longitudinal terms $\sigma_{(\mathrm{L})}(0)$ and $\varpi_{(\mathrm{L})}(0)$. Moreover it shifts the DC Hall terms from their symmetry dictated values.}

{\ With these facts taken care of we are now in a position to make our first important observation. In much of the holographic literature (see \textit{e.g.}~\cite{Amoretti:2018tzw} and references therein), when considering the hydrodynamics of systems where translation invariance is broken solely by scalar operators of the kind discussed above, the effective momentum conservation equation entering the hydrodynamic description was assumed to have the form
    \begin{eqnarray}
      \label{Eq:MomentumConservationwithGamma}
      \partial_{t} P^{i} + \partial_{j} T^{ij} &=& - \Gamma^{ij} P_{j} + F^{i \mu} J_{\mu} - \varphi O_{j} \; , 
    \end{eqnarray}
where $\Gamma^{ij}$ is some putative momentum dissipation tensor. 
Notice however that if we take this equation as a starting point, we can then demonstrate that the correlation functions for a system with a constant $\Gamma^{ij}$ satisfy Ward identities of the form
    \begin{subequations}
        \label{Eq:WardwithGamma}
        \begin{eqnarray}
            \left( i \omega \delta\indices{^{i}_{k}} - \Gamma\indices{^{i}_{k}} \right) \langle Q^{k} Q^{j} \rangle &=& - \left(  \mu \left( i \omega \delta\indices{^{i}_{k}} - \Gamma\indices{^{i}_{k}} \right) - F\indices{^{i}_{k}} \right) \langle Q^{k} J^{j} \rangle + \varphi \langle Q^{i} O^{j} \rangle \nonumber \\
            &\;& - i \omega \left( \chi_{\pi \pi} - \mu n \right) \delta^{ij} \; , \qquad \\
            \left( i \omega \delta\indices{^{i}_{k}} - \Gamma\indices{^{i}_{k}} \right) \langle Q^{k} J^{j} \rangle &=& - \left( \mu \left( i \omega \delta\indices{^{i}_{k}} - \Gamma\indices{^{i}_{k}} \right) - F\indices{^{i}_{k}} \right) \langle J^{k} J^{j} \rangle + \varphi \langle J^{i} O^{j} \rangle \nonumber \\
            &\;& - i \omega n \delta^{ij} \; , \\
            \left( i \omega \delta\indices{^{i}_{k}} - \Gamma\indices{^{i}_{k}} \right) \langle Q^{k} O^{j} \rangle &=& - \left( \mu \left( i \omega \delta\indices{^{i}_{k}} - \Gamma\indices{^{i}_{k}} \right) - F\indices{^{i}_{k}} \right) \langle J^{k} O^{j} \rangle - \varphi \langle O^{i} O^{j} \rangle \nonumber \\
            &\;& + \delta^{ij} \; .
        \end{eqnarray}
    \end{subequations}
These should be compared to \eqref{Eq:ExplicitWardIdentity}. As a direct consequence, for any system coming from \eqref{Eq:ConservationEquations2}, we have 
    \begin{eqnarray}
     \label{Eq:ExplicitHydroIncoherentIdentifications2}
     \Gamma^{ij} = 0 \; .
    \end{eqnarray}
In summary, even if the effective hydrodynamic description of the momentum conservation equation \eqref{Eq:MomentumConservationwithGamma} could somehow have differed from the one-point function Ward identities \eqref{Eq:ConservationEquations2}, for the resultant hydrodynamic correlation functions to satisfy \eqref{Eq:ExplicitWardIdentity} it must be the case that \eqref{Eq:ExplicitHydroIncoherentIdentifications2} holds. 
A nonzero $\Gamma$ features in the hydrodynamic description of several
holographic models in the literature where the breaking of translations is mostly
explicit (\textit{e.g.}~\cite{Andrade:2013gsa,Amoretti:2014zha,Amoretti:2014mma,Amoretti:2015gna,Amoretti:2017xto,Baggioli:2014roa}). In the next section we will show within a concrete hydrodynamic setup
how an effective $\Gamma^{ij}$ emerges when one enters the explicit regime where
the dynamics of the pseudo-Goldstone bosons freeze out.}
 
\section{Simple hydrodynamics of a CDW in a magnetic field}
\label{sec:hydrosec}

{\ Hydrodynamics is an effective approach to the description of interacting systems valid at large distances and late times. It consists of conservation equations and accompanying constitutive relations which express charge currents in terms of the differences in thermodynamic parameters from global equilibrium. Quasihydrodynamics is the partner theory where some of these initially conserved charges are not conserved, but decay. Recent work has shown that the quasihydrodynamic description of certain systems can be valid in regimes where the non-conservation of a hydrodynamic charge is relatively strong \cite{Amoretti_2020} \textit{i.e.}~one where the parameter responsible for breaking the conservation, in our case the magnetic field, does not vanish in the ground state and in fact contributes to the definition of global thermodynamic equilibrium. In what follows we shall apply quasihydrodynamics to the systems discussed above.}

{\ Our model will consist of the typical ingredients for describing the hydrodynamics of a charged fluid in an external magnetic field (see \textit{e.g.}~\cite{Delacretaz:2019wzh}). In the pseudo-spontaneous case we must include two pseudo-Goldstone bosons arising from the breaking of a translational+shift symmetry down to its diagonal subgroup. To start this procedure requires us to supply the form of the currents and the associated conservation equations in global thermodynamic equilibrium. In particular we need to understand how the translation breaking modes contribute to the thermodynamics.}

{\ The free energy ($F$) of our (pseudo-)Goldstone bosons $(\delta O_{i=1,2})$, including linear and quadratic orders in the field, must take the form \cite{Armas:2019sbe,Armas:2020bmo} 
    \begin{eqnarray}
        \label{Eq:Goldstonefreeenergy}
        F &=& \int d^{2}x \; \left[ P_{l}^{0} \left(\partial_{i} \delta O^{i} + \partial_{i} \delta O_{j} \partial^{i} \delta O^{j} \right) + \frac{K}{2} ( \partial_{i} \delta O^{i} )^2 + \frac{G}{2} \left( \partial_{i} \delta O_{j} \partial^{i} O^{j} + k_{0}^2 \delta O_{i} \delta O^{i} \right)  \right.  \qquad \nonumber \\
          &\;& \left. \vphantom{\frac{K}{2}} \hphantom{\int d^{2}x \; \left[ \right.} + \left( \delta P_{l} - \delta s_{\mathrm{tr}} \right) \partial_{i} \delta O^{i} - \delta s_{\mathrm{curl}} \epsilon^{ij} \partial_{i} \delta O_{j} 
                \right] \; , \qquad
    \end{eqnarray}
 where $P_{l}^{0}$ is the background lattice pressure \cite{Armas:2019sbe,Ammon:2020xyv,Armas:2020bmo}, $\delta P_{l}$ its fluctuation with respect to chemical potential and temperature, $K$ is the bulk modulus, $G$ the shear modulus and $k_{0}$ a small mass term (or equivalently a large inverse correlation length) for the boson. We define the ``pinning'' frequency in terms of $k_{0}$ to be
    \begin{eqnarray}
    \label{Eq:Pinningfrequency}
    \omega_{0}^2 &=& \frac{G k_{0}^2}{\chi_{\pi \pi}} \; . 
    \end{eqnarray}
In writing \eqref{Eq:Goldstonefreeenergy} we have made use of spatial rotational invariance, and spatial isotropy for the underlying lattice \cite{Delacretaz:2017zxd,Delacretaz:2019wzh}. We have included $\delta s_{\mathrm{tr}}$ and $\delta s_{\mathrm{curl}}$ as the sources for the (pseudo-)Goldstone bosons, corresponding to the operators $\lambda_{\mathrm{tr}} = \partial_{i} \delta O^{i}$ and $\lambda_{\mathrm{curl}} = \epsilon^{ij} \partial_{i} \delta O_{j}$, that will enter our hydrodynamic description. We choose these operators as they respect rotational symmetry and become invariant under the emergent shift symmetry, $\delta O_{i} \rightarrow \delta O_{i} + a_{i}$, as $k_{0} \rightarrow 0$, \AAm{in analogy with the mechanism described in Section \ref{sec:intro} \cite{Nicolis:2013lma,Nicolis:2015sra}}.}

{\ From the expression for the (pseudo-)Goldstone boson free energy \eqref{Eq:Goldstonefreeenergy} we make two crucial observations. Firstly, we notice that when $k_{0}=0$ the action \eqref{Eq:Goldstonefreeenergy} enjoys both a shift and a spatial translation symmetry. \AAm{Similarly to what has been done in \cite{Delacretaz:2016ivq,Delacretaz:2017zxd,Delacretaz:2019wzh}, we assume that our scalar effectively describes the situation where this joint symmetry is broken to a diagonal subgroup, in such a way that the Goldstone bosons $\delta O_i$ can be considered as the conjugate variables of momentum, satisfying the following Poisson bracket:}
    \begin{eqnarray}
        \label{Eq:Opicorrelator}
         \left\{ \delta O_{i}(t,\vec{x}), P^{j}(t,\vec{y}) \right\}
        &=& \left( \delta\indices{_{i}^{j}} + \partial^{j} \delta O_{i}(t,\vec{x}) \right) \delta^{2}(\vec{x}-\vec{y})\,, 
    \end{eqnarray}
where $P^{i}$ is the generator of spatial translations. Of the two terms on the right hand side the latter is the usual one for translating a vector field by a constant shift. The former however indicates that the bosons have a non-zero ``charge'' under translations. This is important as it subsequently implies non-conservation of the spatial momentum for our model\DKB{. We can see this in the following manner: assume we have access to the full Hamiltonian of the theory and consider the Hamiltonian time evolution equation for the spatial momentum $P^{i}$. Part of this Hamiltonian is given by the contribution to the free energy of the pseudo-Goldstone bosons \eqref{Eq:Goldstonefreeenergy}. Employing \eqref{Eq:Opicorrelator} the Hamiltonian time evolution equation yields}
    \begin{eqnarray}
          \label{Eq:HamiltonianEvolutionBosons}
          \partial_{t} P^{i}
      &=& \left\{ P^{i}, H \right\} 
      = - \omega_0^2 \chi_{\pi \pi} \delta O^{i} + \mathrm{higher \; derivatives} \; .
    \end{eqnarray}
\DKB{As the right hand side is non-zero, the spatial momentum is not a conserved quantity unless the boson becomes massless ($k_{0}=0$).}

{\ Secondly, another consequence of the commutation relations \eqref{Eq:Opicorrelator} is the ``Josephson relation''. This gives the time evolution of the scalar in the Hamiltonian formalism \cite{Son:2000ht,Delacretaz:2017zxd}. In particular, in the presence of a source velocity the Hamiltonian density, to first order in fluctuations, can be taken to have the form $\mathcal{H} = P^{i} v_{i} + \ldots$. Therefore
    \begin{eqnarray}
     \label{Eq:JosephsonRelation}
     \partial_{t} \delta O^{i} &=& \left\{ \delta O^{i}, \mathcal{H} \right\} \stackrel{\eqref{Eq:Opicorrelator}}{=} v^{i}  + \mathrm{higher \; derivatives}.  
    \end{eqnarray}
This expression will enter our hydrodynamic description as the lowest order in derivatives effective equation for the evolution of the scalar operators $\delta O^{i}$. In the effective approach we will have to supplement the right hand side with additional higher derivative terms as one increases the derivative order of hydrodynamics. In the explicit case we must also take account of the fact that the phonons can relax \cite{Delacretaz:2017zxd,Amoretti:2017frz}. This is done through the introduction of a phase relaxation term $\Omega^{ij} \delta O_{j}$ on the right hand side of \eqref{Eq:JosephsonRelation} \textit{i.e.}~
    \begin{eqnarray}
     \label{Eq:JosephsonRelation2}
     \partial_{t} \delta O^{i} &=& - \Omega\indices{^{i}_{j}} \delta O^{j} + v^{i} + \mathrm{higher \; derivative \; terms}.
    \end{eqnarray}
}

{\ On the assumptions we have made in detailing the thermodynamics of our translation breaking scalars, we should ask ourselves what happens as the pinning frequency becomes large. As the pinning frequency increases it will become harder and harder to excite the translation breaking scalars. Of course, there is still a Josephson relation, and the Ward identities will continue to hold as their applicability does not depend on the value of $k_{0}$. Nevertheless, in the range of parameters where the phonon mode is extremely heavy, it becomes non-dynamical. In this case one can neglect the time derivative in the Josephson relation \eqref{Eq:JosephsonRelation2} which subsequently constraints $\delta O^i$ in terms of the fluid velocity $v^i$. Substituting the solution to this constraint into the equation of motion for the momentum density \eqref{Eq:HamiltonianEvolutionBosons} one finds:
    \begin{equation}\label{Eq:effectiveGammahydro}
        \partial_t P^i=-\frac{\omega_0^2}{ \chi_{\pi \pi}} \left(\Omega^{-1} \right)^i_{\; j} P^j+... \ ,
    \end{equation}
where the ellipsis represents terms not relevant for this discussion. Equation \eqref{Eq:effectiveGammahydro} is exactly the equation of motion one should expect for the momentum density in a system which exhibits explicit momentum relaxation if one identifies $\Gamma^i_j\equiv\omega_0^2\,\left(\Omega^{-1} \right)^i_{\, j}$. This shows that in the purely explicit limit (meaning when $k_0$ is large), one recovers (as expected) an effective description in terms of a standard momentum dissipation rate $\Gamma$, with the latter being expressed in terms of the ``old'' phonon variables $\omega_0$ and $\Omega\indices{^{i}_{j}}$. The correlator will look like the standard correlator of charged hydrodynamics with an external magnetic field and momentum dissipation \cite{Sachdev}.}

{\ The observations above, \eqref{Eq:WardwithGamma} and \eqref{Eq:effectiveGammahydro}, apply regardless of the specifics of the hydrodynamic system. From this point onward we shall make a simplifying assumption - namely, we will neglect the lattice pressure $P_{l}^{0}$ and its thermodynamic derivatives $\delta P_{l}$ in \eqref{Eq:Goldstonefreeenergy}. In fact $P_{l}^{0}$ can be set to zero if one restricts the analysis to thermodynamically stable vacua \cite{Ammon:2020xyv,Armas:2020bmo}. Thermodynamic derivatives of the lattice pressure, $\delta P_{l}$, might be non-zero even in these configurations but they do not appear in the set of correlators that we consider in this work \cite{Armas:2020bmo}. We shall then show how one can apply the Ward identities to the hydrodynamic correlators to evaluate the hydrodynamic transport coefficients that appear in the constitutive relations.} 

{\ With these assumptions at hand, we find that the contribution of $\delta O_{i}$ to the spatial stress tensor, obtained by applying the Noether procedure on the symmetry $\delta O_{i} \rightarrow \delta O_{i} + a_{i}$, has the form
    \begin{eqnarray}
      \label{Eq:NoetherSEM}
      T_{ij}^{O} &=& - \left( K \partial_{k} \delta O^{k} \delta_{ij} + G \partial_{i} \delta O_{j} \right) \; , 
    \end{eqnarray}
when the source terms vanish. This expression can be made symmetric while preserving the translational Ward identity, as expected for a theory which has spatial rotation invariance, by subtracting the term $\Delta T_{ij}^{O} = G \left( \partial_{j} \delta O_{i} - \delta_{ij} \partial_{k} \delta O^{k} \right)$ which is divergence free in the ``$i$'' index. The spatial part of the stress tensor can also be re-expressed in terms of $\lambda_{\mathrm{tr}} = \partial_{i} \delta O^{i}$ and $\lambda_{\mathrm{curl}} = \epsilon^{ij} \partial_{i} \delta O_{j}$ at the cost of leaving an explicitly anti-symmetric term in the expression. This is achieved by adding $\Delta T_{ij}^{O} $ to \eqref{Eq:NoetherSEM},
    \begin{eqnarray}
     \label{Eq:SEMcontributionofphi}
     T_{ij}^{O} &=& - \left( \left( K + G \right) \lambda_{\mathrm{tr}} \delta_{ij} + G \lambda_{\mathrm{curl}} \epsilon_{ij} \right) \; . 
    \end{eqnarray}
The latter expression \eqref{Eq:SEMcontributionofphi} will appear at leading order in our constitutive relation for the spatial stress tensor of the full theory ($T^{ij}$).}

{\ We can also derive the static susceptibility for our theory. It has the form
    \begin{eqnarray}
        \label{Eq:StaticSusceptibility}
     \chi &=& \left(
        \begin{array}{cccccc}
          \partial_{T} s & \partial_{\mu} s & 0 & 0 & 0 & 0 \\
          \partial_{T} n & \partial_{\mu} n & 0 & 0 & 0 & 0 \\
          0 & 0 & \chi_{\pi \pi} & 0 & 0 & 0 \\
          0 & 0 & 0 & \chi_{\pi \pi} & 0 & 0 \\
          0 & 0 & 0 & 0 & \chi_{\mathrm{tr}} = \frac{\vec{k}^2}{G k_{0}^2 + (K + G) \vec{k}^2 } & 0 \\
          0 & 0 & 0 & 0 & 0 & \chi_{\mathrm{curl}} = \frac{\vec{k}^2}{G ( k_{0}^2 + \vec{k}^2 )} \\
        \end{array}
     \right) \; ,
    \end{eqnarray}
where $s$ is the entropy density and we work in the basis of sources $s_{A}=(T, \mu, \vec{v}, s_{a})$ with $\vec{v}$ the spatial velocity of the fluid. To derive the susceptibilities for the Goldstone bosons $\partial_{s_{b}} \lambda_{a}$ one needs simply rewrite the static equations of motion following from \eqref{Eq:Goldstonefreeenergy} including the displayed non-zero source terms. We recall that if one considers the lattice pressure to be non-zero (non-thermodynamically stable vacua), additional terms would appear in \eqref{Eq:StaticSusceptibility} \cite{Armas:2019sbe,Ammon:2020xyv,Armas:2020bmo}. These terms do not influence the results obtained in this paper.

\subsection{Spontaneous case}

{\ With the susceptibilities~\eqref{Eq:StaticSusceptibility} to hand we are in a position to begin building the hydrodynamic constitutive relations. We start in the spontaneous case, where there is no source term for the scalars in the background. The constitutive relations we derive are to first order in derivatives and fluctuations of temperature $T+\delta T$, chemical potential $\mu +\delta \mu$ and scalar sources $\delta s_{\mathrm{tr}}$ and $\delta s_{\mathrm{curl}}$ assuming that the spatial velocity $v^{i}$ vanishes in equilibrium.}

\subsubsection{Constitutive relations}

{\ To lay out our notation for the transport coefficients it is easiest to first consider the constitutive relations for the fluctuations of the vector currents corresponding to heat and electric charge flow which take the form
    \begin{eqnarray}
     \label{Eq:Heatconstitutive}
     \delta Q^{i} &=& \left( \chi_{\pi \pi} - \mu n \right) v^{i} + \alpha^{ij} \left( \delta E_{j} + F_{jk}  v^{k} - \partial_{j} \delta \mu \right) 
                      - \kappa^{ij} \frac{\partial_{j} \delta T}{T} \nonumber \\
                  &\;&  - \theta^{ij} \partial_{j} \delta s_{\mathrm{tr}} - \iota^{ij} \epsilon\indices{_{j}^{k}} \partial_{k} \delta s_{\mathrm{curl}} \; , \\
     \label{Eq:Chargeconstitutive}
     \delta J^{i} &=& n  v^{i} + \sigma^{ij} \left( \delta E_{j} + F_{jk}  v^{k} - \partial_{j} \delta \mu \right) - \alpha^{ij} \frac{\partial_{j} \delta T}{T} \nonumber \\
                &\;& - \gamma^{ij} \partial_{j} \delta s_{\mathrm{tr}} + \gamma^{ij} \epsilon\indices{_{j}^{k}} \partial_{k} \delta  s_{\mathrm{curl}} \; , \qquad \\
     F^{ij} &=& B \epsilon^{ij} \; , \qquad \epsilon^{12} = 1 \; , \nonumber 
    \end{eqnarray}
to order one in derivatives and fluctuations where $\delta E^{i}$ is an electric field fluctuation and $\epsilon^{ij}$ is the two dimensional Levi-Civita tensor. The tensor transport coefficients can be decomposed with respect to $SO(2)$ rotational and microscopic spatial parity invariance \textit{e.g.}~
    \begin{eqnarray}
        (\sigma,\alpha,\kappa,\gamma,\theta,\iota,\ldots)^{ij} &=& (\sigma,\alpha,\kappa,\gamma,\theta,\iota,\ldots)_{\mathrm{(L)}} \delta^{ij}  
                                                            + (\sigma,\alpha,\kappa,\gamma,\theta,\iota,\ldots)_{\mathrm{(H)}} F^{ij} \; . \qquad
    \end{eqnarray}
The above decomposition of the transport coefficients appearing in the hydrodynamic constitutive relations is with respect to $F_{ij}$ and not its inverse; unlike the AC transport coefficients \eqref{Eq:Spontaneousdecomp}. This reflects the fact that these terms should be smooth as we take $B \rightarrow 0$. We also note that while permitted by the gradient expansion, $\iota_{(\mathrm{L})}$ and $\iota_{(\mathrm{H})}$ shall not appear in the diffusive sector of the spontaneous case (this will be different in the explicit case where their value will be fixed).}

{\ With these definitions to hand we can also write down the Josephson condition to first order in derivatives and fluctuations (starting from \eqref{Eq:JosephsonRelation}). Upon also imposing the Onsager conditions we find
    \begin{eqnarray}
          \label{Eq:JosephsonSpontaneousConstitutive}
     &\;& \partial_{t} \delta O^{i} + J^{i}_{O} =0 \; , \nonumber \\
     J^{i}_{O} &=& - v^{i} + \gamma^{ij} \left( \delta E_{j} + F_{jk} v^{k} - \partial_{j} \delta \mu \right) \nonumber \\
                     &\;& - \left( \theta_{(\mathrm{L})} \delta^{ij} - \iota_{(\mathrm{H})} F^{ij} \right) \frac{\partial_{j} \delta T}{T}
                         - \chi_{\mathrm{tr}} \left( \xi_{\mathrm{tr},(\mathrm{L})} \delta^{ij} + \xi_{\mathrm{tr},(\mathrm{H})} F^{ij} \right) \partial_{j} \delta s_{\mathrm{tr}} \nonumber \\
                      &\;& - \chi_{\mathrm{curl}} \left( \xi_{\mathrm{curl},(\mathrm{L})} \delta^{ij} + \xi_{\mathrm{curl},(\mathrm{H})} F^{ij} \right) \epsilon\indices{_{j}^{k}} \partial_{k} \delta s_{\mathrm{curl}} \; , 
    \end{eqnarray}
 where we have employed spatial rotation invariance to decompose the terms proportional to derivatives of the scalar sources into longitudinal and Hall parts. This has been done to make clear constraints imposed by the Onsager relations. In \eqref{Eq:JosephsonSpontaneousConstitutive} we have, a little erroneously, treated $\xi_{\mathrm{tr},(\mathrm{H})}$ and $\xi_{\mathrm{curl},(\mathrm{H})}$ as distinct variables. From the Onsager relations however one finds
    \begin{eqnarray}
        \label{Eq:XHdef}
        X_{(\mathrm{H})} \equiv \frac{\xi_{\mathrm{tr},(\mathrm{H})}}{G + K} = - \frac{\xi_{\mathrm{curl},(\mathrm{H})}}{G} \; .
    \end{eqnarray}
Similarly, in \cite{Amoretti_2019} spatial rotation invariance implies
    \begin{eqnarray}
        \label{Eq:XLdef}
        X_{(\mathrm{L})} \equiv \frac{\xi_{\mathrm{tr},(\mathrm{L})}}{G + K} = - \frac{\xi_{\mathrm{curl},(\mathrm{L})}}{G} \; . 
    \end{eqnarray}
We shall henceforth collect these transport coefficients into a new tensor structure $X^{ij}$.}

{\ Last, but not least, we must supply the constitutive relation for the fluctuation of the spatial stress tensor. This has the form
    \begin{eqnarray}
            \label{Eq:Spatialstresstensor}
            \delta T_{ij} 
        &=& \left( n \delta \mu + \left( \chi_{\pi \pi} - \mu n \right) \frac{\delta T}{T}- ( G + K ) \chi_{\mathrm{tr}} \delta s_{\mathrm{tr}} \right) \delta_{ij} - G \chi_{\mathrm{curl}} \delta s_{\mathrm{curl}} \epsilon_{ij} \nonumber \\
        &\;& - 2 \eta \sigma_{ij} - \zeta \partial_{k} v^{k} \delta_{ij} \; , \\
            \sigma_{ij}
        &=& \frac{1}{2} \left( \partial_{i} v_{j} + \partial_{j} v_{i} \right) - \partial_{k} v^{k} \delta_{ij} \; ,
    \end{eqnarray}
The first two terms in \eqref{Eq:Spatialstresstensor} are the standard thermodynamic contributions to the stress tensor from charge and heat. The second two terms are contributions from the Goldstone boson \eqref{Eq:SEMcontributionofphi}. The second line of terms in \eqref{Eq:Spatialstresstensor} are all first order in derivatives and standard for a relativistic fluid.}

\subsubsection{The AC correlators}

{\ With the constitutive relations to hand we can employ the Martin-Kadanoff procedure to determine the finite frequency response of the system. Of particular interest are the relations that result when we substitute these expressions into the Ward identities. The hydrodynamic correlators must satisfy the Ward identities at arbitrary frequency. From this substitution we immediately learn, by expanding the resultant expressions at large frequency and zero wavevector, that
    \begin{eqnarray}
     \label{Eq:Transportcoeffconstraints}
     \alpha^{ij} = - \mu \sigma^{ij} \; , \qquad \kappa^{ij} = \mu^2 \sigma^{ij} \; , \qquad \theta^{ij} = - \mu \gamma^{ij} \; . 
    \end{eqnarray}
These constraints are the usual relations one finds for a system which has Lorentz covariance (even if the ground state of the theory breaks Lorentz invariance) given our choice of heat current and normalisations of the transport coefficients. The coefficients $\iota_{(\mathrm{L})}$ and $\iota_{(\mathrm{H})}$ remain unconstrained. Fortunately neither of these coefficients enter into the diffusive sector for spontaneous breaking. In the explicit case, where they do enter, we will fix them using the Ward identities.}

{\ The relativistic constraints on the hydrodynamic transport coefficients \eqref{Eq:Transportcoeffconstraints} can readily be seen from the Ward identities at large frequency. At low frequency we can use the Ward identities of \eqref{Eq:LowFrequencySpontaneousWard} and the AC hydrodynamic correlators to determine the hydrodynamic transport coefficients in terms of the DC values of those same correlators. \DKB{Of particular interest are the incoherent electric charge conductivities (or the transport coefficients $\sigma_{(\mathrm{L})}$ and $\sigma_{(\mathrm{H})}$). In the absence of a magnetic field these parameterise the flow of charge orthogonal to the transport of spatial momentum and are independent of the momentum loss rate. In the presence of a magnetic field this clean definition does not hold because the flow of spatial momentum is constantly twisting direction due to its coupling to $B$. Instead, following \cite{Amoretti_2020}, we take a naive definition as the constant term in the Laurent expansion of the electric conductivities about the diffusive pole at zero wavevector. This definition is invariant under the replacement $\omega \rightarrow \omega - i \Gamma$, with $\Gamma$ the momentum loss rate, mirroring a key property of the zero magnetic field incoherent conductivities (independence of momentum loss).} }

{\ We find for the incoherent electric charge conductivities
    \begin{subequations}
        \begin{eqnarray}
            \Xi\, \sigma_{(\mathrm{L})} 
        &=& \chi_{\pi \pi}^2 \kappa_{(\mathrm{L})}(0) \; , \\
            \Xi\, \sigma_{(\mathrm{H})} 
        &=& \frac{n}{B^2}  \left(B^2 \kappa_{(\mathrm{L})}(0)^2+ \left( \kappa_{(\mathrm{H})}(0) + \mu ^2 n \right)^2 \right) - \frac{2 \mu  \chi _{\pi \pi }^3}{B^2}  \nonumber \\
        &\;& + \frac{\chi _{\pi \pi }^2}{B^2} \left(\kappa_{(\mathrm{H})}(0)+5 \mu ^2 n \right) - \frac{4 \mu  n  \chi _{\pi \pi}}{B^2} \left(\kappa_{(\mathrm{H})}(0)+\mu ^2 n \right)
        \; , \\
            \label{Eq:DefinitionofXi}
            \Xi
        &\equiv& B^2 \kappa_{(\mathrm{L})}(0)^2+\left(\kappa_{(\mathrm{H})}(0)+\mu ^2 n -2 \mu  \chi _{\pi \pi }\right){}^2 \; ,
    \end{eqnarray}
    \end{subequations}
where the lack of argument on $\sigma_{(\mathrm{L})}$ and $\sigma_{(\mathrm{H})}$ indicates that these are transport coefficients of the constitutive relations (\textit{e.g.}~\eqref{Eq:Heatconstitutive}), while $\kappa_{(\mathrm{L})}(0)$ and $\kappa_{(\mathrm{H})}(0)$ are DC values for the thermal conductivities. Additionally we have expressions for the hydrodynamic transport coefficients $\gamma_{(\mathrm{L})}$, $\gamma_{(\mathrm{H})}$, $X_{(\mathrm{L})}$ and $X_{(\mathrm{H})}$ which we shall present in a compactified form shortly (see \eqref{Eq:SpontaneousIncoherentConductivities}). As per magneto-transport without broken translation invariance we see that in the spontaneous case, generically, there is a non-zero incoherent Hall conductivity.}

{\ Having imposed all the constraints implied by the Ward identities we can now understand the response of our system to time varying fields. For example, the electric conductivities at non-zero frequency in the spontaneous case have almost exactly the same form as those for the dyonic black hole \cite{Sachdev,Hartnoll_ohm,Amoretti_2020} in terms of the momentum susceptibility $\chi_{\pi \pi}$ and the DC thermal conductivities. In particular
    \begin{eqnarray}
     \label{Eq:ACspontaneoussigmas}
     \sigma_{(\mathrm{L})}(\omega) &=& \frac{i \omega  \chi_{\pi \pi} \left( \gamma_{\mathrm{c}}^2 - i \gamma_{\mathrm{c}}\, \omega + \omega_{\mathrm{c}}^2\right)}
                                            {B^2 \left( ( \omega + i \gamma_{\mathrm{c}})^2 - \omega_{\mathrm{c}}^2 \right)} \; , \qquad \\
     \sigma_{(\mathrm{H})}(\omega) &=& - \left( n + \frac{ \chi _{\pi \pi } \omega^2 \omega_{\mathrm{c}}}{B \left( ( \omega + i \gamma_{\mathrm{c}})^2 - \omega_{\mathrm{c}}^2 \right)} \right) \; ,
    \end{eqnarray}
where the cyclotron frequency and decay rate are
    \begin{eqnarray}
     \omega_{\mathrm{c}} = \frac{B \chi_{\pi \pi} \left( \mu ^2 n - 2 \mu  \chi _{\pi \pi } + \kappa_{\mathrm{(H)}}(0) \right)}
                             {\Xi} \, , \qquad 
     \gamma_{\mathrm{c}} = \frac{B^2 \chi _{\pi \pi } \kappa_{\mathrm{(L)}}(0)}{\Xi} \; , 
    \end{eqnarray}
 respectively. The longitudinal conductivity expressed in terms of the cyclotron frequency and decay rate has the same form as found in \cite{Hartnoll_ohm}, the key difference between \cite{Amoretti_2020} and \cite{Hartnoll_ohm} being a different identification of the cyclotron frequency and cyclotron decay rate. The Hall conductivity differs between the two because in \cite{Amoretti_2020} there is no simple expression for $n$ in terms of $\omega_{\mathrm{c}}$.}

{\ As was noted previously, it is only necessary to know $\langle J^{i} J^{j} \rangle$ and $\langle J^{i} O^{j} \rangle$ at arbitrary frequency to determine all other correlators that appear in the Ward identities. For this latter correlator, its value at non-zero frequency is
    \begin{eqnarray}
        \label{Eq:ACspontaneousgammas}
        \gamma_{(\mathrm{L})}(\omega) &=& \frac{\omega  \left( B \omega  \omega_{\mathrm{c}} \theta_{(\mathrm{L})}(0) - \left(\gamma_{\mathrm{c}} (\omega +i \gamma_{\mathrm{c}})+i \omega_{\mathrm{c}}^2\right) \left(\theta_{(\mathrm{H})}(0) +\mu \right)\right)}{B^2 \left( (\omega + i \gamma_{\mathrm{c}})^2 - \omega_{\mathrm{c}}^2 \right)} \; , \\
        \gamma_{(\mathrm{H})}(\omega) &=& \frac{\omega ^2 \omega_{\mathrm{c}} \left(\theta_{(\mathrm{H})}(0)+\mu \right) + B \left( i \omega  \left(\gamma_{\mathrm{c}}^2 - i \gamma_{\mathrm{c}} \omega + \omega_{\mathrm{c}}^2\right) \theta_{(\mathrm{L})}(0) + (\omega + i \gamma_{\mathrm{c}} )^2 - \omega_{\mathrm{c}}^2\right)}{B \left( (\omega + i \gamma_{\mathrm{c}})^2 - \omega_{\mathrm{c}}^2 \right)} \; , \nonumber \\
    \end{eqnarray}
where $\theta_{(\mathrm{L})}(0)$ and $\theta_{(\mathrm{H})}(0)$ are the longitudinal and Hall components of the zero frequency limit of $\langle Q^{i} O^{j} \rangle$. We remind the reader that this is the system dependent information necessary to specify the zero frequency behaviour of any system satisfying our assumptions.}

{\ The AC correlator involving two Goldstone bosons is separated from the other correlators as it does not enter into the Ward identities (unlike in the explicit case). Consequently, its values are not generally constrained by the Ward identities and DC data. Instead, we can only learn information about this correlator by specifying that we are in the hydrodynamic regime. The expression is somewhat complex so it is worthwhile to introduce some new notation. Let
    \begin{eqnarray}
     \hat{\sigma} = \sigma_{(\mathrm{L})} \mathbbm{1}_{2} - \sigma_{(\mathrm{H})} F \; , \; \;
     \hat{\gamma} = \gamma_{(\mathrm{L})} \mathbbm{1}_{2} - \gamma_{(\mathrm{H})} F \; , \; \; \hat{\Gamma} = F \cdot \hat{n} \; , \; \;
     \hat{I} = \mathbbm{1}_{2} + F \cdot \hat{\gamma}  \; , \qquad \nonumber \\
          \hat{n} = \frac{1}{\chi_{\pi \pi}} \left( n \mathbbm{1}_{2} - F \cdot \hat{\sigma} \right) 
          \; , \qquad
          (\hat{\kappa},\hat{\theta})(0) = (\kappa,\theta)_{(\mathrm{L})}(0) \mathbbm{1}_{2}
     - (\kappa,\theta)_{(\mathrm{H})}(0) F^{-1} \; , \qquad 
    \end{eqnarray}
then the Goldstone-Goldstone correlator takes the form
        \begin{eqnarray}
            \label{Eq:SpontaneousGoldstoneCorrelators}
            \hat{X}(\omega) = \hat{X} - \frac{i \omega}{\chi_{\pi \pi}} \hat{I} \cdot \hat{I} \cdot \hat{\Lambda}^{-1}(\omega) \; , \qquad
            \hat{\Lambda}(\omega) = - i \omega \mathbbm{1}_{2} \cdot \left( - i \omega \mathbbm{1}_{2} + \hat{\Gamma} \right) \; .
        \end{eqnarray}
Simultaneously, we can rewrite our other correlators in this unified notation (which will be particularly useful for the explicit case). As such, the independent transport coefficients appearing in the hydrodynamic constitutive relations are
    \begin{subequations}
    \label{Eq:SpontaneousIncoherentConductivities}
    \begin{eqnarray}
     \hat{\sigma} &=& \hat{\Phi}^{-1} \left( n \hat{\kappa}(0) - \left( \mu n - \chi_{\pi \pi} \right)^{2} F^{-1} \right) \; , \\
     \hat{\gamma} &=& \hat{\Phi}^{-1} \left( \mu \left( \mu n - \chi_{\pi \pi} \right) F^{-1} - \chi_{\pi \pi} \hat{\theta}(0) - \hat{\kappa}(0) \right) \; , \\
     \hat{X} &=& \hat{\Phi}^{-1} \left( \mu \left( \mu n - 2 \chi_{\pi \pi} \right) \hat{X}(0) - \mu^2 F^{-1} - \hat{X}(0) \cdot F \cdot \hat{\kappa}(0) \right. \nonumber \\
     &\;& \left. \hphantom{\hat{\Phi}^{-1} \left( \right.} + 2 \mu \hat{\theta}(0) - \hat{\theta}(0) \cdot F \cdot \hat{\theta}(0) \right) \; , \qquad \\
     \hat{\Phi} &=& F \cdot \hat{\kappa}(0) - \left( \mu^2 n - 2 \mu \chi_{\pi \pi} \right) \mathbbm{1}_{2} \; ,
    \end{eqnarray}
    \end{subequations}
which allows us to rewrite \eqref{Eq:ACspontaneoussigmas} and \eqref{Eq:ACspontaneousgammas} as
    \begin{subequations}
    \begin{eqnarray}
     \hat{\sigma}(\omega) &=& \hat{\sigma} - i \omega \chi_{\pi \pi} \hat{n} \cdot \hat{n} \cdot \hat{\Lambda}^{-1}(\omega) \; , \\
     \hat{\gamma}(\omega) &=& \hat{\gamma} + i \omega \hat{n} \cdot \hat{I} \cdot  \hat{\Lambda}^{-1}(\omega) \; ,
    \end{eqnarray}
    \end{subequations}
with $\hat{\Lambda}(\omega)$ given in \eqref{Eq:SpontaneousGoldstoneCorrelators}.}

\subsection{Explicit case}

Distinctly from the spontaneous case, and recalling that due to \eqref{Eq:ExplicitHydroIncoherentIdentifications2} there is no momentum relaxation rate ($\Gamma^{ij}=0$), in the explicit case there is the potential for a non-zero phase relaxation term $\Omega^{ij}$ in the Josephson relation, namely\footnote{One can instead assume the most general thing possible (\textit{i.e.}~$\Gamma^{ij} \neq 0$) for the equations of motion, so that
    \begin{eqnarray}
      \partial_{t} P^{i} + \partial_{j} T^{ij} &=& - \Gamma^{ij} P_{j} - n E^{i} + F^{ij} J_{j} - \omega_0^2 \chi_{\pi \pi} O^i \; .
    \end{eqnarray}
However, the comparison of the previous equations against the Ward identities \eqref{Eq:ExplicitWardIdentity} at large frequency implies $\Gamma^{ij}=0$, in accordance with \eqref{Eq:ExplicitHydroIncoherentIdentifications2}.
} 
    \begin{eqnarray}
      \partial_{t}  O^{i} + J^{i}_{O} &=& - \Omega\indices{^{i}_{j}}  O^{j} \; , \label{Eq:Josephsonexplicit} \\
      \partial_{t} P^{i} + \partial_{j} T^{ij} &=& - n E^{i} + F^{ij} J_{j} - \omega_0^2 \chi_{\pi \pi} O^i \; . \label{Eq:eomphydro}
    \end{eqnarray}
We can employ $SO(2)$ spatial rotation invariance and microscopic spatial parity invariance to decompose the phase relaxation tensor into
    \begin{eqnarray}
        \Omega^{ij} = \Omega_{(\mathrm{L})} \delta^{ij} + \Omega_{(\mathrm{H})} F^{ij} \; ,
    \end{eqnarray}
where the coefficients must be determined from data.
    
{\ Modulo these modifications to the (non-)conservation equations, the situation is not so different from the spontaneous case. In particular the constitutive relations are unmodified. However, to employ our formalism we shall have to assume that the pinning frequency, and consequently $k_{0}$, is sufficiently small such that the dynamics of the scalar modes are of the pseudo-Goldstone type and hence they are long-lived enough to enter our effective hydrodynamic description. This can be seen post-hoc by comparing AC correlators to real data.}

\subsubsection{The AC correlators}

{\ We can once again employ the Martin-Kadanoff procedure and determine the constraints that result when we substitute the correlators into the Ward identities. First, by comparing the 1-pt Ward identity \eqref{Eq:ConservationEquations2} with \eqref{Eq:eomphydro} we find
    \begin{eqnarray}
        \varphi = \omega_{0}^2 \chi_{\pi \pi} \, ,
    \end{eqnarray}
which fixes the source of the scalar $\varphi$ in terms of the pinning frequency, or using \eqref{Eq:Pinningfrequency} in terms of the mass of the phonon $k_0^2$. Additionally, from expanding the Ward identities at large frequencies we learn that
    \begin{eqnarray}
     \label{Eq:ExplicitHydroIncoherentIdentifications}
      \iota^{ij} = \mu \gamma^{ij} \,,
    \end{eqnarray}
which constrain transport coefficients that were unconstrained in the spontaneous case.}

{\ Given these new identifications we find that the independent AC correlators, from which all others can be derived using \eqref{Eq:ExplicitWardIdentity}, are 
    \begin{subequations}
    \label{Eq:ExplicitACcorrelators}
    \begin{eqnarray}
        \hat{\sigma}(\omega) &=& \hat{\sigma} + \left( \chi_{\pi \pi} \hat{n} \cdot \hat{n} \cdot \left( - i \omega \mathbbm{1}_{2} + \hat{\Omega} \right) \right. \nonumber \\
        &\;& \left. \hphantom{\hat{\sigma} + \left( \right.}
            + \chi_{\pi \pi}\omega_{0}^2 \left( \hat{n} \cdot \hat{\gamma} \cdot \left( 2 \mathbbm{1}_{2} + F \cdot \hat{\gamma} \right) + i \omega \hat{\gamma} \cdot \hat{\gamma} \right) \right)  \hat{\Xi}^{-1}(\omega) \; , \\
        \hat{\varpi}(\omega) &=& \left( \hat{n} \cdot \hat{I} + \left( i \omega \mathbbm{1}_{2} - \hat{\Gamma} \right) \cdot \hat{\gamma} \right) \hat{\Xi}^{-1}(\omega) \; , \\
         \hat{\zeta}(\omega) &=& \frac{1}{\omega_{0}^2 \chi_{\pi \pi}} \left( i \omega \mathbbm{1}_{2} - \hat{\Gamma} \right)  \hat{\Xi}^{-1}(\omega) \; , \\
              \hat{\Xi}(\omega) &=& \left( - i \omega \mathbbm{1}_{2} + \hat{\Gamma} \right) \left( - i \omega \mathbbm{1}_{2} + \hat{\Omega} \right) + \omega_{0}^2 \hat{I} \cdot \hat{I} \; ,
    \end{eqnarray}
    \end{subequations}
where
     \begin{eqnarray}
     \hat{\Omega} &=& \Omega_{(\mathrm{L})} \mathbbm{1}_{2} - \Omega_{(\mathrm{H})} F \; .
    \end{eqnarray}
To arrive at \eqref{Eq:ExplicitACcorrelators} we have imposed the constraints \eqref{Eq:Transportcoeffconstraints}, \eqref{Eq:ExplicitHydroIncoherentIdentifications} and \eqref{Eq:ExplicitHydroIncoherentIdentifications2}.}

 Again, we can extract the incoherent conductivities (and other hydrodynamic transport coefficients) in terms of the DC conductivities of our system using the low frequency Ward identities. One point of note is that, in opposition to what is found for the spontaneous case, the electric and thermo-electric DC conductivities contain information about the system beyond what is dictated by symmetry. We can therefore use them to constrain the hydrodynamic transport coefficients entering the constitutive relations entirely in terms of the DC electric, thermo-electric and thermal conductivities. For example, the independent transport coefficients appearing in the hydrodynamic constitutive relations for the currents are
    \begin{subequations}
    \begin{eqnarray}
     \sigma_{(\mathrm{L})} &=& - \frac{1}{2} \mathrm{Tr}\left[\hat{\Psi}^{-1} \left( n \left( \hat{\kappa}(0) \cdot F \cdot \hat{\sigma}(0) - \hat{\alpha}(0) \cdot F \cdot \hat{\alpha}(0) \right) - n^2 \hat{\kappa}(0) \right. \right. \nonumber \\
     &\;& \left. \left. \hphantom{\hat{\Psi}^{-1} \left( \right.} + 2 n \left( \chi_{\pi \pi} - \mu n \right) \hat{\alpha}(0) - \left( \chi_{\pi \pi} - \mu n\right)^2 \hat{\sigma}(0) \right) \right] \; , \\
     \sigma_{(\mathrm{H})} &=& \frac{1}{2} \mathrm{Tr}\left[F^{-1} \hat{\Psi}^{-1} \left( n \left( \hat{\kappa}(0) \cdot F \cdot \hat{\sigma}(0) - \hat{\alpha}(0) \cdot F \cdot \hat{\alpha}(0) \right) - n^2 \hat{\kappa}(0) \right. \right. \nonumber \\
     &\;& \left. \left. \hphantom{\hat{\Psi}^{-1} \left( \right.} + 2 n \left( \chi_{\pi \pi} - \mu n \right) \hat{\alpha}(0) - \left( \chi_{\pi \pi} - \mu n\right)^2 \hat{\sigma}(0) \right) \right] \; , \\
     \gamma_{(\mathrm{L})} &=& - \frac{1}{2} \mathrm{Tr} \left[ \hat{\Psi}^{-1} \left( n \hat{\kappa}(0) + \mu \left( \mu n - \chi_{\pi \pi} \right) \hat{\sigma}(0) + \left( 2 \mu n - \chi_{\pi \pi} \right) \hat{\alpha}(0) \right. \right. \nonumber \\
     &\;& \left. \left. \hphantom{\hat{\Psi}^{-1} \left( \right.} - \hat{\kappa}(0) \cdot F \cdot \hat{\sigma}(0) + \hat{\alpha}(0) \cdot F \cdot \hat{\alpha}(0) \right) \right] \; , \\
     \gamma_{(\mathrm{H})} &=& \frac{1}{2} \mathrm{Tr} \left[ F^{-1} \hat{\Psi}^{-1} \left( n \hat{\kappa}(0) + \mu \left( \mu n - \chi_{\pi \pi} \right) \hat{\sigma}(0) + \left( 2 \mu n - \chi_{\pi \pi} \right) \hat{\alpha}(0) \right. \right. \nonumber \\
     &\;& \left. \left. \hphantom{\hat{\Psi}^{-1} \left( \right.} - \hat{\kappa}(0) \cdot F \cdot \hat{\sigma}(0) + \hat{\alpha}(0) \cdot F \cdot \hat{\alpha}(0) \right) \right] \; , \\
     \hat{\Psi} &=& \chi_{\pi \pi}^2 \mathbbm{1}_{2} + \mu \left( \mu n - 2 \chi_{\pi \pi} \right) F \cdot \hat{\sigma}(0) + 2 \left( \mu n - \chi_{\pi \pi} \right) F \cdot \hat{\alpha}(0) \nonumber \\
     &\;& + n F \cdot \hat{\kappa}(0) + ( F \cdot \hat{\alpha}(0) )^2 - F \cdot \hat{\kappa}(0) \cdot F \cdot \hat{\sigma}(0)\; , \\
     (\hat{\sigma},\hat{\alpha},\hat{\kappa})(0) &=& (\sigma,\alpha,\kappa)_{(\mathrm{L})}(0) \mathbbm{1}_{2}
     + (\sigma,\alpha,\kappa)_{(\mathrm{H})}(0) F \;  .
    \end{eqnarray}
    \end{subequations}
Notice the redefinition of $\hat{\kappa}$ compared to the spontaneous case. These expressions in terms of the thermo-electric conductivities are the most convenient for comparison against any putative experiment, as they do not require any specification of the DC values of correlators involving the scalar. However, the interested reader can use the Ward identities \eqref{Eq:Explicitwardlowfrequency} to switch out dependence on $\sigma_{(\mathrm{L}),(\mathrm{H})}$ and $\alpha_{(\mathrm{L}),(\mathrm{H})}$ for $\varpi_{(\mathrm{L}),(\mathrm{H})}$ and $\zeta_{(\mathrm{L}),(\mathrm{H})}$ if they so choose.

Finally, to completely specify the correlators in the explicit case we need expressions for the phase relaxation rates. When performing the above identification procedure for the hydrodynamic transport coefficients one additionally finds
    \begin{subequations}
    \label{Eq:Omegaidentification}
    \begin{eqnarray}
     \Omega_{(\mathrm{L})} &=& \frac{\omega_{0}^2 \chi_{\pi \pi}}{2} \mathrm{Tr}\left[\hat{\Psi}^{-1} \left( \hat{\kappa}(0) + 2 \mu \hat{\alpha}(0) + \mu^2 \hat{\sigma}(0) \right) \right] \,,  \\
     \Omega_{(\mathrm{H})} &=& - \frac{\omega_{0}^2 \chi_{\pi \pi}}{2} \mathrm{Tr}\left[F^{-1} \hat{\Psi}^{-1} \left( \hat{\kappa}(0) + 2 \mu \hat{\alpha}(0) + \mu^2 \hat{\sigma}(0) \right) \right] \, .
    \end{eqnarray}
    \end{subequations}
These expressions are valid at arbitrary magnetic fields and pinning frequencies, including the $B \rightarrow 0$ limit with $\omega_{0}>0$, so long as we remain within the hydrodynamic regime. Importantly, the limits of the trace terms in \eqref{Eq:Omegaidentification} as $\omega_{0} \rightarrow 0$ are smooth and non-zero for any $B>0$, and thus the leading dependence on $\omega_{0}$ of the phase relaxation coefficients at small pinning frequency is $\omega_{0}^2$. This becomes important when comparing the explicit scalar-scalar correlator as $\omega_{0} \rightarrow 0$ to the spontaneous correlator. The exact value of the proportionality coefficient can be computed by examining the scalar-scalar correlator of the explicit case in such a limit and requesting that it match the spontaneous correlator \cite{Amoretti:2018tzw,Baggioli:2020nay,Baggioli:2020haa}. Consequently we obtain
    \begin{eqnarray}
     \label{Eq:PhaseRelaxationtoSpontaneousIdentification}
     \Omega_{(\mathrm{L})} = \chi_{\pi \pi} \omega_{0}^{2} X_{(\mathrm{L})} + \mathcal{O}(\omega_{0}^4)  \, , \qquad \Omega_{(\mathrm{H})}= \chi_{\pi \pi} \omega_{0}^{2} X_{(\mathrm{H})} + \mathcal{O}(\omega_{0}^4) \, , 
    \end{eqnarray}
for arbitrary values of the magnetic field. The first identification was already discussed in \cite{Amoretti:2018tzw,Donos:2019hpp,Baggioli:2020nay,Baggioli:2020haa}. Here we have generalized this relation to include $\Omega_{(\mathrm{H})}$ (which is present only at non-zero magnetic field) and most importantly we have avoided any ordering ambiguities compared to the $B=0$ case as in that situation $\omega_{0} \rightarrow 0$ and $k \rightarrow 0$ do not commute as limits.

\section{Summary and future work}
\label{sec:conclusions}

{\ In this paper we have developed a formalism for computing the diffusion characteristics of $(2+1)$-dimensional charged fluids in an external magnetic field with broken translation invariance.
We have worked at order one in derivatives and fluctuations and considered 
a breaking mechanism where a scalar operator acquires a spatially modulated vev. We observed that the system does not include any fundamental momentum dissipation effects except as an effective description when the dynamics of the translation breaking scalars are frozen out. We have supplied analytic expressions for the hydrodynamic transport coefficients appearing in the constitutive relations in terms of experimentally measurable quantities: the DC thermo-electric transport coefficients.}

{\ It would be interesting to use our formalism to more precisely understand the behaviour of certain holographic models with (pseudo-)spontaneous breaking of translation invariance \cite{Andrade:2017cnc,Alberte:2017oqx,Amoretti:2017frz,Amoretti:2017axe,Alberte:2017cch,Amoretti:2018tzw,Andrade:2018gqk,Amoretti:2019kuf,Amoretti_2019,Ammon:2020xyv,Amoretti:2020ica,Andrade:2020hpu,Baggioli:2020nay,Baggioli:2020edn}. This necessarily will require us to consider non-zero lattice pressures (as these systems are unstable) \cite{Armas:2019sbe,Ammon:2020xyv,Armas:2020bmo}. The resultant hydrodynamics is more complicated but we do not expect any significant divergences from what we have found here and are currently pursuing this study. In this context, it would also be worthwhile to perform a more systematic scan of the parameter space of these holographic models in an effort to elucidate the bounds on our hydrodynamic approach, particularly in light of current programmes examining the convergence radius of hydrodynamics \cite{Withers_2018,Grozdanov:2019kge,Heller:2020uuy}.}

{\ We might also consider modifying our action to include terms that can generate the effective actions described in \cite{Delacretaz:2019wzh} \textit{i.e.}~time dependencies for the scalar of the form $\epsilon^{ij} O_{i} \partial_{t} O_{j}$. Indeed, as discussed in \cite{Delacretaz:2019wzh}, there should be a limit of our analytic expressions where the magnetoplasmon is pushed out of the hydrodynamic regime. What is not clear to us is whether, when  we take this limit, the effective action describing the scalar fluctuations in our system includes the desired kinetic term. We are also exploring this issue.}

{\ The applications of the approach of \cite{Amoretti_2020} are only in the initial phases of exploration. A fundamental question to address in this direction is to apply the method to models which realise translation symmetry breaking in a different way from the class analysed here, \textit{e.g.}~via a spatially modulated charge density profile, and study if in this case the correct hydrodynamic theory is still the one described in this work.}

{\ An additional potential extension is to consider generalising our formalism to $(3+1)$-dimensions and tackling anomalous transport. A comprehensive approach to such systems without momentum dissipation and assuming a vanishing spatial velocity in the ground state is discussed in recent work \cite{Ammon:2020rvg}. However it is expected that these anomalous fluids will generally have a non-zero velocity in the ``laboratory frame''. This has presented several problems in the literature as without momentum dissipation this system is inherently unstable. Additionally, in formulations that incorporate a non-zero velocity in the ground state, there are known problems in satisfying the Onsager relations \cite{Abbasi:2018zoc}. }

\section*{Acknowledgments}
We would like to thank Blaise Gout\'eraux for carefully reading an earlier version of the manuscript. We also thank Luca Martinoia for having found several typos in a previous version of the manuscript. D. A. is supported by the `Atracci\'on de Talento' programme
(2017-T1/TIC-5258, Comunidad de Madrid) and through the grants SEV-2016-0597 and PGC2018-095976-B-C21.
This work has been partially supported by the INFN Scientific Initiative SFT: “Statistical Field Theory, Low-Dimensional Systems, Integrable Models and Applications”.
\appendix

\bibliography{references}
\bibliographystyle{JHEP}

\end{document}